\documentclass[pdflatex,sn-mathphys-num]{sn-jnl}% Math and Physical Sciences Numbered Reference Style
\usepackage{graphicx}%
\usepackage{multirow}%
\usepackage{amsmath,amssymb,amsfonts}%
\usepackage{amsthm}%
\usepackage{mathrsfs}%
\usepackage[title]{appendix}%
\usepackage{xcolor}%
\usepackage{textcomp}%
\usepackage{manyfoot}%
\usepackage{booktabs}%
\usepackage{algorithm}%
\usepackage{algorithmicx}%
\usepackage{algpseudocode}%
\usepackage{listings}%
\theoremstyle{thmstyleone}%
\theoremstyle{thmstyletwo}%

\theoremstyle{thmstylethree}%

\begin{document}

\title[Approximate Likelihood-Based Inference for Spatial Generalized Linear Mixed Models]{Approximate Likelihood-Based Inference for Spatial Generalized Linear Mixed Models}


\author*[1]{\fnm{Samuel I.} \sur{Watson}}\email{s.i.watson@bham.ac.uk}

\author[1]{\fnm{Yixin} \sur{Wang}}\email{y.wang19@bham.ac.uk}

\author[1]{\fnm{Emanuele} \sur{Giorgi}}\email{e.giorgi@bham.ac.uk}

\affil*[1]{\orgdiv{Department of Applied Health Sciences}, \orgname{University of Birmingham}, \orgaddress{ \city{Birmingham}, \postcode{B152TT}, \state{West Midlands}, \country{United Kingdom}}}


\abstract{We study maximum likelihood estimation for spatial generalized linear mixed models with Gaussian process approximations using a stochastic Newton–Raphson algorithm. We consider two Gaussian Process approximations in this context: spectral Gaussian process approximations and stochastic partial differential equations (SPDE). We refine the stochastic maximum likelihood algorithm and we propose a new stopping criterion for efficient termination to prevent long runs of sampling in the stationary post-convergence phase and a Monte Carlo estimator of fixed effect standard errors. We run a series of simulation comparisons of spatial statistical models alongside the popular Bayesian integrated nested Laplacian approximation method which incorporates SPDE. We show that HSGP provides nominal coverage of fixed and random effect parameters with `smooth' latent fields but performance degrades for rough fields. SPDE in a stochastic maximum likelihood framework maintains nominal coverage and matches or improves upon the performance of Bayesian integrated nested Laplacian approximation.}




\maketitle

\section{Introduction}\label{sec-intro}
Generalised linear mixed models (GLMM) are a highly flexible class of statistical models that incorporate both `fixed' or population-level effects and `random' or group- or subject-specific effects. GLMMs permit the incorporation of latent effects and parameters and allow for complex covariance structures. For example, they are widely used in the analysis of: clustered data \citep{Li2021}, cohort studies, and in geospatial statistical models as the realisation of a Gaussian process used to model a latent spatial or spatio-temporal surface \citep{Diggle1998}, which is the topic of our study.

A common approach to estimation of GLMMs in a Frequentist setting is to approximate the integrand using a Laplace approximation. This method is employed by popular software including \textit{lme4} for R \citep{lme4}, and by Stata's xtmixed function. The quality of the Laplace approximation can degrade in certain circumstances. \citet{Shun1995}, \citet{Ogden2021}, and others demonstrate that the error of the Laplace approximation increases when the dimension of the integral (here, in the log likelihood) is not sufficiently large relative to the sample size. A key example where this may not hold is in spatial or spatio-temporal mixed models, where the model's random effects represent the realisation of a Gaussian process \citep{Diggle1998}, and where the dimension of the integral is equivalent to the sample size.  

An alternative approach is Monte Carlo Maximum Likelihood (MCML). \citet{mcculloch1997maximum} described three variants of these algorithms, which are expectation-maximisation algorithms involving maximising the expectation of the log-likelihood components by averaging over Monte Carlo samples of the random effects. Typically, Markov Chain Monte Carlo (MCMC) is used to generate the samples. Subsequent developments have considered spatial models (e.g. \citet{Zhang2002, Cheng2022}) and other non-linear models (e.g. \citep{Meza2009}). \citet{Jank2006} described a Stochastic Approximation Expectation Maximisation (SAEM) algorithm, which is a Ruppert-Monroe algorithm, that is a potentially more efficient approach to MCML as it reuses random effect samples between iterations. Some recent work has considered incorporation of covariance matrix approximations into these algorithms \citep{Guan2021} and other authors have proposed incorporation of restricted maxmium likelihood estimators in the context of SAEM \citep{Meza2009} or more generally using a Monte Carlo scheme \citep{Liao2002}. We use a stochastic Newton-Raphson algorithm for estimation.

Full maximum likelihood estimation of a Gaussian Process model scales as $O(n^3)$ due to needing to invert the covariance matrix of the latent field or random effects. The poor scaling of Gaussian process models has driven a focus on approximate inference \citep{Simpson2012}, however most methods concentrate on Bayesian paradigm. Many of these approximations may also be relevant in a maximum likelihood context as well, but have received little attention in this context. We study the `Hilbert Space Gaussian Process' (HSGP) \citep{Solin2020} and stochastic partial differential equation (SPDE) \citep{Lindgren2011} approximations and incorporate them into our stochastic Newton-Raphson algorithm. 

In a spatial setting, approximate Bayesian estimation has become popular, including integrated nested laplacian approximation (INLA) \citep{Rue2009}. INLA is an approximation to the Bayesian posterior, but may also have good Frequentist characteristics. INLA implements the SPDE approximation to the Gaussian Process. As a popular tool it provides a benchmark for the comparisons in this study. Other approximations to the covariance matrix of a Gaussian Process are typically framed in a Bayesian context  \citep{Solin2020, Datta2016, RiutortMayol2023}. The lack of focus on maximum likelihood may be due to the complexity of estimation algorithms in this context. Nevertheless, there may be a desire for reliable maximum likelihood methods with spatial models, such as in experimental contexts and estimation of treatment effects in a geographic setting \citep{Watson2025}.

In this article, we investigate likelihood-based estimation of the spatial mixed model using HSGP and SPDE. We compare the performance and running times of our algorithms with INLA. One limitation of stochastic algorithms is knowing when to stop. \citet{Caffo2005} proposed stopping criteria for MCML algorithms, but these potentially allow long phases of iterations post-convergence. We propose a stopping criterion to reduce running times. We point out though that the focus of this paper is limited to assessing the performance of the proposed algorithm and INLA-based point and interval estimation procedures from a data-driven perspective. In this context, the use of Bayesian inference methods should be seen primarily as a computational device and comparator. Our comparison is valid insofar as non or weakly-informative priors are used in the INLA implementation, and the goal of the comparison is to compare Frequentist charactersitcs of the estimators. Under these conditions, both approaches aim to extract similar information from the data, making performance comparisons meaningful and practically relevant for applied researchers choosing between estimation frameworks.

\section{Statistical Methods}
\subsection{Generalised Linear Mixed Models}
We consider generalised linear mixed models (GLMM) with the linear predictor for observation $i$
\begin{equation*}
    \eta_i = \mathbf{x}_i\boldsymbol{\beta} + \mathbf{z}_i \mathbf{u}
\end{equation*}
where $\mathbf{x}_i$ is the $i$th row of matrix $X$, which is a $n \times P$ matrix of covariates, $\boldsymbol{\beta}$ is a vector of parameters. In addition, $\mathbf{z}_i$ is the $i$th row of matrix $Z$, which is the $n \times Q$ ``design matrix'' for the random effects, and $\mathbf{u} \sim N(0,D)$, where $D$ is the $Q \times Q$ covariance matrix of the random effects terms that depends on parameters $\boldsymbol{\theta}$. In this article we consider that $D$ is the realisation of a stationary, isotropic Gaussian process so that the $(i,i')th$ elements of $D$ are $D_{ii'} = C(s_i,s_{i'}) = \tau^2 k(||s_i - s_{i'}||; \lambda)$ where $||.||$ is the Euclidean distance, $\lambda$ is a length-scale parameter, $\tau^2$ is the variance of the Gaussian process, and $k(.)$ is a correlation function. To simplify the exposition we assume $Z = I$ without loss of generality, hence $Q = n$ and we write $\mathbf{u} = u(\mathbf{s})$ where required to make explicit the location of the observations.

The model for $i$ is then
\begin{equation*}
y_i \sim G(h(\eta_i))
\end{equation*}
$\mathbf{y}$ is a $n$-length vector of outcomes with elements $y_i$, $G$ is an exponential-family distribution, $h(.)$ is the link function such that $\mu_i = h(\eta_i)$ where $\mu_i$ is the mean value. We only consider non-linear link functions here as Gaussian-linear models do not require the Monte Carlo fitting method described in the article.

The likelihood of this model is given by:
\begin{equation}
\label{eq:lik1}
    L(\beta,\theta|\mathbf{y}) = \int \prod_{i=1}^n f_{y|\mathbf{u}}(y_i|\mathbf{u}, \beta)f_{\mathbf{u}}(\mathbf{u}|\theta) d \mathbf{u}
\end{equation}
where $f_{\mathbf{u}}$ is the multivariate Gaussian denisty function. Where relevant we represent the set of all parameters as $\boldsymbol{\Theta} = (\boldsymbol{\beta}, \boldsymbol{\theta})$.

\subsection{Hilbert Space Gaussian Process Approximation}
\label{sec:hsgp}

The computational bottleneck in fitting spatial GLMMs is the $O(n^3)$ cost of operations involving the dense $n \times n$ covariance matrix $D$. The Hilbert Space Gaussian Process (HSGP) approximation \citep{Solin2020,RiutortMayol2023} replaces $D$ with a low-rank diagonal representation derived from a spectral decomposition of the covariance function, reducing the effective dimension of the random effects from $n$ to $M \ll n$.

\subsubsection{Spectral Approximation}

Consider our stationary covariance function $C(s, s')$ on a bounded domain $[-L_d, L_d]^D$. By Bochner's theorem, the covariance function admits a spectral density $S(\boldsymbol{\omega})$. The HSGP approximation represents the Gaussian process as
\begin{equation*}
    u(\mathbf{s}) \approx \sum_{j=1}^{M} v_j \phi_j(\mathbf{s})
\end{equation*}
where $\phi_j(\mathbf{s})$ are the Laplacian eigenfunctions on the domain:
\begin{equation*}
    \phi_j(\mathbf{s}) = \prod_{d=1}^{D} \frac{1}{\sqrt{L_d}} \sin\left(\frac{m_{jd} \pi (s_d + L_d)}{2L_d}\right)
\end{equation*}
with multi-index $\mathbf{m}_j = (m_{j1}, \ldots, m_{jD})$, and $v_j$ are independent spectral coefficients with prior
\begin{equation*}
    v_j \sim N(0, \Lambda_j), \qquad \Lambda_j = S(\boldsymbol{\omega}_j)
\end{equation*}
where $\boldsymbol{\omega}_j$ has components $\omega_{jd} = m_{jd}\pi / (2L_d)$ and $S$ is the spectral density evaluated at these frequencies. The total number of basis functions is $M = \prod_{d=1}^D m_d$. For the Mat\'{e}rn class with smoothness $\nu$:
\begin{equation}
\label{eq:spd_matern}
    S(\boldsymbol{\omega}) = \sigma^2 \frac{(4\pi)^{D/2}\,\Gamma(\nu + D/2)}{\Gamma(\nu)} \frac{(2\nu)^\nu}{\ell^{2\nu}} \left(\frac{2\nu}{\ell^2} + \|\boldsymbol{\omega}\|^2\right)^{-(\nu + D/2)}
\end{equation}

\subsubsection{GLMM Formulation}

Within the GLMM framework of Section 2.1, the HSGP approximation corresponds to replacing the dense covariance model with an equivalent model having design matrix $\Phi$ and diagonal covariance. Let $\Phi$ denote the $n \times M$ matrix with entries $\Phi_{ij} = \phi_j(\mathbf{s}_i)$. The linear predictor becomes
\begin{equation*}
    \eta_i = \mathbf{x}_i \boldsymbol{\beta} + \boldsymbol{\phi}_i \mathbf{u}, \qquad \mathbf{u} \sim N(\mathbf{0}, \text{diag}(\boldsymbol{\Lambda}))
\end{equation*}
where $\boldsymbol{\phi}_i$ is the $i$th row of $\Phi$. In the notation of Section 2.1, the effective random effects design matrix becomes $\Phi$ and the covariance matrix $D = \text{diag}(\boldsymbol{\Lambda})$. The marginal likelihood retains the same form as (\ref{eq:lik1}) with these substitutions, and all operations involving $D$ reduce to $O(M)$:
\begin{equation*}
    D^{-1} = \text{diag}(1/\boldsymbol{\Lambda}), \qquad
    \log\det(D) = \sum_{j=1}^M \log \Lambda_j, \qquad
    \mathbf{u}^T D^{-1} \mathbf{u} = \sum_{j=1}^M u_j^2 / \Lambda_j
\end{equation*}

\subsection{Stochastic Partial Differential Equation Approximation}
\label{sec:spde}
An alternative sparse approach to avoiding the $O(n^3)$ cost of operations involving the dense covariance matrix $D$ is the stochastic partial differential equation (SPDE) approximation of \citet{Lindgren2011}, which underpins the spatial models fit by INLA. Whereas HSGP replaces $D$ with a low-rank diagonal representation, SPDE retains the full-dimensional random effect vector but works with a sparse precision matrix $Q = D^{-1}$ derived from a finite-element discretisation. Operations with $Q$ scale as $O(n_v^{3/2})$ in two dimensions, where $n_v$ is the number of mesh vertices.

\subsubsection{SPDE Representation}
A Gaussian process with Mat\'{e}rn covariance on $\mathbb{R}^d$ is the stationary solution of the stochastic partial differential equation
\begin{equation*}
    (\kappa^2 - \Delta)^{\alpha/2} u(\mathbf{s}) = \mathcal{W}(\mathbf{s})
\end{equation*}
where $\mathcal{W}$ is Gaussian white noise, $\Delta$ is the Laplacian, and the smoothness parameter is related to the order of the SPDE by $\nu = \alpha - d/2$ \citep{Lindgren2011}. We consider $d = 2$ and $\nu = 1$ to match the simulation setting of Section 4, which corresponds to $\alpha = 2$.

Approximate solutions are constructed using a finite element basis $\{\psi_j\}_{j=1}^{n_v}$ of piecewise-linear tent functions over a triangular mesh covering the domain, so that
\begin{equation*}
    u(\mathbf{s}) \approx \sum_{j=1}^{n_v} u_j \psi_j(\mathbf{s})
\end{equation*}
The coefficient vector $\mathbf{u}$ has a multivariate Gaussian distribution with a sparse precision matrix:
\begin{equation}
\label{eq:spde_Q}
    Q(\sigma^2, \lambda) = a_C C + a_G G + a_M M
\end{equation}
where $C$ is the lumped mass matrix, a diagonal matrix with entries $C_{jj} = \int \psi_j(\mathbf{s})\,d\mathbf{s}$; $G$ is the stiffness matrix with entries $G_{jk} = \int \nabla \psi_j(\mathbf{s})^T \nabla \psi_k(\mathbf{s})\,d\mathbf{s}$; and $M = G C^{-1} G$. With the reparameterisation $\kappa = 2\sqrt{2}/\lambda$ and $\tau^2 = 1/(4\pi \kappa^2 \sigma^2)$, the coefficients are
\begin{equation*}
    a_C = \frac{2}{\pi \sigma^2 \lambda^2}, \qquad a_G = \frac{1}{2\pi \sigma^2}, \qquad a_M = \frac{\lambda^2}{32 \pi \sigma^2}
\end{equation*}
The mesh-dependent matrices $C$, $G$, and $M$ do not depend on the covariance parameters and can be precomputed once; $Q$ inherits their shared sparsity pattern.

\subsubsection{GLMM Formulation}
The field is evaluated at observation locations through a sparse barycentric projector $A$, an $n \times n_v$ matrix with three non-zero entries per row corresponding to the vertices of the triangle containing the location of observation $i$. Within the GLMM framework of Section 2.1, the linear predictor becomes
\begin{equation*}
    \eta_i = \mathbf{x}_i \boldsymbol{\beta} + (Z A \mathbf{u})_i, \qquad \mathbf{u} \sim N(\mathbf{0}, Q^{-1})
\end{equation*}
where $Z$ retains its original role of mapping observations to spatial locations and reduces to the identity for purely geospatial designs. The effective random effects design matrix is therefore $Z_A := Z A$, and the covariance matrix is $D = Q^{-1}$. Unlike HSGP, $D$ is dense and is never formed explicitly; all operations are carried out through the sparse precision $Q$. A sparse Cholesky factorisation $Q = L_Q L_Q^T$, computed with an approximate minimum degree fill-reducing permutation, yields $O(n_v^{3/2})$ complexity in two dimensions \citep{Rue2005}, and gives
\begin{equation*}
    \log \det(D) = -2 \sum_{j=1}^{n_v} \log (L_Q)_{jj}, \qquad \mathbf{u}^T D^{-1} \mathbf{u} = \mathbf{u}^T Q \mathbf{u}
\end{equation*}
The quadratic form is computed directly from the sparse $Q$ without a linear solve.

\subsection{Maximum Likelihood Estimation}
Our main model fitting approach is Monte Carlo Maximum Likelihood (MCML). MCML algorithms have three steps \citep{mcculloch1997maximum}, where on the $t$th step:
\begin{enumerate}
    \item Generate a sample of values of the `random effects' $\hat{\mathbf{u}}^{(t)}$ conditional on the data and current values of $\hat{\beta}^{(t-1)}$ and $\hat{\theta}^{(t-1)}$.
    \item Update the estimates of $\hat{\beta}^{(t)}$, averaging over the sample of $\hat{\mathbf{u}}^{(t)}$, conditional on the current values $\hat{\theta}^{(t)}$ and the data.
    \item Update the estimates of $\hat{\theta}^{(t)}$, averaging over the sample of $\hat{\mathbf{u}}^{(t)}$, conditional on the values of $\hat{\beta}^{(t)}$ and the data.
\end{enumerate}
These steps are then repeated until a convergence criterion is met. We discuss each of these steps in turn. Our approach relies on stochastic Newton-Raphson steps for each of the three steps of the algorithm. We discuss Monte Carlo sample sizes and stopping criteria in the following section. Our main discussion concerns the implementation of HSGP GLMM models but an equivalent description for full maximum likelihood is given in the supplementary information.

\subsubsection{Sampling of Random Effects}
Previous discussions of these algorithms have generally considered Markov Chain Monte Carlo (MCMC) based sampling of the random effects. While recent advances in MCMC samplers have reduced the time per effective sample size, they may still be computationally intensive, especially if a larger number of independent samples is required, and they require burn-in periods for each step. As an alternative, we consider an importance weighting scheme using the Laplace Gaussian approximation to the posterior distribution of the random effects as a proposal distribution ((e.g. \citet{McCulloch1994,Fellner1986,Schall1991,breslow1993approximate}), which was also described by \citet{Kuk1999}.

The posterior mode of the spectral coefficients is obtained by iteratively reweighted least squares. The posterior precision is
\begin{equation}
\label{eq:hsgp_precision}
    P = \Phi^T W \Phi + \text{diag}(1/\boldsymbol{\Lambda})
\end{equation}
which is an $M \times M$ matrix and where $W = \text{diag}\left( \left(\frac{\partial h^{-1}(\boldsymbol{\eta})}{\partial \boldsymbol{\eta}}\right)^2 \text{Var}(\mathbf{y}| \mathbf{u})\right)^{-1}$, which are recognisable as the GLM iterated weights \citep{breslow1993approximate, mccullagh2019generalized}. Samples are drawn from the Gaussian proposal $\mathbf{u}^{(k)} \sim N(\bar{\mathbf{u}}, P^{-1})$ and reweighted by importance sampling. The importance weights are
\begin{equation*}
    w_k^* \propto \exp\left(\log f_{\mathbf{y}|\mathbf{u}}(\mathbf{y}|\mathbf{u}^{(k)}, \boldsymbol{\beta}) - \frac{1}{2}\sum_{j=1}^M \log\Lambda_j - \frac{1}{2}\sum_{j=1}^M \frac{u_j^{(k)2}}{\Lambda_j} - \log q(\mathbf{u}^{(k)})\right)
\end{equation*}
where $q$ denotes the Gaussian proposal density. Both the Cholesky factorisation for sampling and the IRLS iterations operate on $M \times M$ matrices.

The posterior mode of the mesh coefficients is obtained by iteratively reweighted least squares. The posterior precision is
\begin{equation}
\label{eq:spde_precision}
    P = Z_A^T W Z_A + Q
\end{equation}
which is an $n_v \times n_v$ sparse matrix and where $W$ is the diagonal matrix of GLM iterated weights defined as in Section \ref{sec:hsgp}. Both the prior precision $Q$ and the observation term $Z_A^T W Z_A$ are sparse, so $P$ inherits a sparse structure and admits an efficient Cholesky factorisation $P = L_P L_P^T$ under a fill-reducing permutation. Samples are drawn from the Gaussian proposal $\mathbf{u}^{(k)} \sim N(\bar{\mathbf{u}}, P^{-1})$ by computing $\mathbf{u}^{(k)} = \bar{\mathbf{u}} + L_P^{-T} \mathbf{z}_k$ with $\mathbf{z}_k \sim N(\mathbf{0}, I)$, which requires only a sparse triangular solve per sample. The importance weights are
\begin{equation*}
    w_k^* \propto \exp\left(\log f_{\mathbf{y}|\mathbf{u}}(\mathbf{y}|\mathbf{u}^{(k)}, \boldsymbol{\beta}) + \sum_{j=1}^{n_v} \log (L_Q)_{jj} - \frac{1}{2}\mathbf{u}^{(k)T} Q \mathbf{u}^{(k)} - \log q(\mathbf{u}^{(k)})\right)
\end{equation*}
where $q$ denotes the Gaussian proposal density and $L_Q$ is the Cholesky factor of the prior precision $Q$. Unlike the HSGP case, the matrices $P$ and $Q$ are of dimension $n_v \times n_v$ rather than $M \times M$, but sparsity ensures the Cholesky factorisations and triangular solves scale as $O(n_v^{3/2})$ rather than cubically.

\subsubsection{Fitting the fixed effect parameters}
We update $\beta$ conditional on $\mathbf{y}$ and $\mathbf{u}^{(t)}$. The approach is independent of any approximation as it is conditional on the random effects. \citet{mcculloch1997maximum} suggested the Newton-Raphson step for maximimising the expectation of the negative log-likelihood over the random effects $E_u\left[ - \log f_{\mathbf{y}|\mathbf{u}}(\mathbf{y}|\mathbf{u},\beta,\phi) \right]$:
\begin{equation*}
        \beta^{(t+1)} = \beta^{(t)} + E_u \left[ X^T W(\beta^{(t)},\mathbf{u}) X \right] ^{-1}X^T \left( E_u \left[ W(\beta^{(t)},\mathbf{u}) \frac{\partial h^{-1}(\boldsymbol{\eta})}{\partial \boldsymbol{\eta}} (\mathbf{y - \boldsymbol{\mu}}(\beta^{(t)},\mathbf{u}))|\mathbf{y} \right] \right)
\end{equation*}
where we make explicit the dependence of matrices on model parameters. For many of the spatial models we investigate, there is often strong correlation between the random effects and the intercept and other fixed effects leading to bias in $\beta$. As such, we mean-centre the random effect samples in the optimisation so that $\boldsymbol{\mu}^*(\beta^{(t)},\mathbf{u}^{(k,t)}) = h(X\beta^{(t)} + \mathbf{u}^{(k,t)} - \bar{\mathbf{u}}^{(k,t)})$. We then estimate the gradient as:
\begin{equation*}
    E_u \left[ W(\beta^{(t)},\mathbf{u}) \frac{\partial h^{-1}(\boldsymbol{\eta})}{\partial \boldsymbol{\eta}} (\mathbf{y} - \boldsymbol{\mu}^*(\beta^{(t)},\mathbf{u}))|\mathbf{y} \right] \approx \sum_{k = 1}^{m_t} w_k^{(t)} W(\beta^{(t)},\mathbf{u}^{(k,t)}) \frac{\partial h^{-1}(\boldsymbol{\eta})}{\partial \boldsymbol{\eta}} (\mathbf{y} - \boldsymbol{\mu}^*(\beta^{(t)},\mathbf{u}^{(k,t)}))
\end{equation*}
and similarly for the inverse Hessian matrix.

\subsubsection{Covariance parameter estimation}
The final step of each iteration is to generate new estimates of $\boldsymbol{\theta}$ given the samples of the random effects by minimising $E_{\mathbf{u}} \left[ -\log(f_{\mathbf{u}}(\mathbf{u}|\theta)) \right]$. For HSGP, the covariance parameters $\boldsymbol{\theta} = (\log\sigma^2, \log\ell)$ enter only through $\boldsymbol{\Lambda}(\boldsymbol{\theta})$. The diagonal structure of $D$ simplifies the gradient and Hessian from matrix traces to sums over the $M$ basis functions. The gradient is
\begin{equation}
\label{eq:hsgp_grad}
    \frac{\partial \log f_{\mathbf{u}}}{\partial \theta_j} = -\frac{1}{2}\sum_{k=1}^M \frac{1}{\Lambda_k}\frac{\partial \Lambda_k}{\partial \theta_j} + \frac{1}{2}\sum_{k=1}^M \frac{u_k^2}{\Lambda_k^2}\frac{\partial \Lambda_k}{\partial \theta_j}
\end{equation}
where the first term is deterministic and the second is estimated by averaging over the Monte Carlo samples. The expected Hessian has elements
\begin{equation}
\label{eq:hsgp_hess_prior}
    M_{\theta_{jl}} = \frac{1}{2}\sum_{k=1}^M \frac{1}{\Lambda_k^2}\frac{\partial \Lambda_k}{\partial \theta_j}\frac{\partial \Lambda_k}{\partial \theta_l}
\end{equation}
For the HSGP the ratio $(\partial \Lambda_k / \partial \log\sigma^2)/\Lambda_k = 1$ for all $k$, which can lead to poor conditioning of $M_\theta$ when the ratios $(\partial \Lambda_k / \partial \log\ell)/\Lambda_k$ are similar across basis functions. To improve conditioning, we supplement (\ref{eq:hsgp_hess_prior}) with the observation-space Fisher information:
\begin{equation}
\label{eq:hsgp_hess_obs}
    H_{jl}^{\text{obs}} = \sum_{i} w_i \left(\frac{\partial \boldsymbol{\eta}}{\partial \theta_j}\right)^T W \left(\frac{\partial \boldsymbol{\eta}}{\partial \theta_l}\right)
\end{equation}
where $\partial\boldsymbol{\eta}/\partial\theta_j = \Phi\,\text{diag}\left(\partial\sqrt{\boldsymbol{\Lambda}}/\partial\theta_j\right)\mathbf{u}$, and the total Hessian used for the Newton-Raphson step is $M_\theta + H^{\text{obs}}$.

The derivatives of the spectral density with respect to the log-transformed parameters are available in closed form. For $\theta_1 = \log\sigma^2$ we have $\partial \Lambda_k / \partial \log\sigma^2 = \Lambda_k$, and for $\theta_2 = \log\ell$ with the Mat\'{e}rn covariance:
\begin{equation*}
    \frac{\partial \Lambda_k}{\partial \log\ell} = \Lambda_k\left(-2\nu + \frac{4\nu(\nu + D/2)}{\ell^2\left(2\nu/\ell^2 + \|\boldsymbol{\omega}_k\|^2\right)}\right)
\end{equation*}

For SPDE, the covariance parameters $\boldsymbol{\theta} = (\log\sigma^2, \log\lambda)$ enter through the precision matrix $Q(\boldsymbol{\theta})$ defined in (\ref{eq:spde_Q}). Because $Q$ is linear in the coefficients $(a_C, a_G, a_M)$ and these depend on $\boldsymbol{\theta}$ only through elementary functions, the derivatives are available in closed form:
\begin{equation}
\label{eq:spde_dQ}
    \frac{\partial Q}{\partial \log \sigma^2} = -Q, \qquad \frac{\partial Q}{\partial \log \lambda} = -2 a_C C + 2 a_M M
\end{equation}
Both derivatives inherit the sparsity pattern of $Q$. Since $\log f_{\mathbf{u}}(\mathbf{u}|\theta) = \tfrac{1}{2}\log\det Q - \tfrac{1}{2}\mathbf{u}^T Q \mathbf{u} + \text{const}$, the gradient with respect to $\theta_j$ is
\begin{equation}
\label{eq:spde_grad}
    \frac{\partial \log f_{\mathbf{u}}}{\partial \theta_j} = \frac{1}{2}\text{tr}\left(Q^{-1}\frac{\partial Q}{\partial \theta_j}\right) - \frac{1}{2}\mathbf{u}^T \frac{\partial Q}{\partial \theta_j} \mathbf{u}
\end{equation}
where the first term is deterministic and the second is estimated by averaging the quadratic form over the Monte Carlo samples. The identity $\partial Q / \partial \log\sigma^2 = -Q$ yields the closed-form result $\text{tr}(Q^{-1}\,\partial Q/\partial\log\sigma^2) = -n_v$, so the $\log\sigma^2$ gradient reduces to $-\tfrac{1}{2}n_v + \tfrac{1}{2}E_{\mathbf{u}}[\mathbf{u}^T Q \mathbf{u}]$. The trace for $\log\lambda$ has no closed form and is estimated stochastically, as described in Section \ref{sec:hutchinson}.

The expected Hessian has elements
\begin{equation}
\label{eq:spde_hess_prior}
    M_{\theta_{jl}} = \frac{1}{2}\text{tr}\left(Q^{-1}\frac{\partial Q}{\partial \theta_j} Q^{-1} \frac{\partial Q}{\partial \theta_l}\right)
\end{equation}
Substituting (\ref{eq:spde_dQ}) gives analytic simplifications for three of its four entries:
\begin{align*}
    M_{\log\sigma^2, \log\sigma^2} &= \frac{n_v}{2}, \qquad
    M_{\log\sigma^2, \log\lambda} = -\frac{1}{2}\text{tr}\left(Q^{-1}\frac{\partial Q}{\partial \log\lambda}\right) \\
    M_{\log\lambda, \log\lambda} &= \frac{1}{2}\text{tr}\left(Q^{-1}\frac{\partial Q}{\partial \log\lambda} Q^{-1} \frac{\partial Q}{\partial \log\lambda}\right)
\end{align*}
The cross-term reuses the trace already computed for the $\log\lambda$ gradient, so only $M_{\log\lambda, \log\lambda}$ requires additional computation. Unlike the HSGP case, the $\log\sigma^2$ direction is separated from $\log\lambda$ by the distinct algebraic structure of $\partial Q/\partial\log\sigma^2$ and $\partial Q/\partial\log\lambda$, and we do not require the observation-space supplement of (\ref{eq:hsgp_hess_obs}). The total Hessian used in the Newton-Raphson step is $M_\theta$ alone.

\subsubsection{Stochastic trace estimation}
\label{sec:hutchinson}
The gradient (\ref{eq:spde_grad}) and Hessian (\ref{eq:spde_hess_prior}) involve traces of the form $\text{tr}(Q^{-1} A)$ and $\text{tr}(Q^{-1} A Q^{-1} A)$ with $A = \partial Q/\partial \log\lambda$. Forming $Q^{-1}$ explicitly would destroy sparsity and defeat the purpose of the SPDE representation. We instead use Hutchinson's stochastic trace estimator \citep{Hutchinson1990}: for any square matrix $B$ and independent probe vectors $\mathbf{z}_r \sim N(\mathbf{0}, I_{n_v})$, $r = 1, \ldots, R$,
\begin{equation}
\label{eq:hutchinson}
    \text{tr}(B) \approx \frac{1}{R}\sum_{r=1}^R \mathbf{z}_r^T B \mathbf{z}_r
\end{equation}
with variance decreasing as $O(1/R)$ \citep{Avron2011}. Both traces are evaluated by applying $B$ to each probe vector through sparse triangular solves against the cached Cholesky factor of $Q$: the gradient trace requires one solve per probe, and the Hessian trace requires two. We use $R = 50$ probes by default.

The trace estimator introduces additional Monte Carlo noise into the Newton-Raphson step for $\boldsymbol{\theta}$, on top of the noise already present from averaging over the sampled $\mathbf{u}^{(k,t)}$. Two modifications are used to control its impact. First, we reuse the same probe vectors $\{\mathbf{z}_r\}$ across the gradient and Hessian traces within a single outer iteration, a common random number scheme \citep{Glasserman1993} that correlates the errors in numerator and denominator of the Newton-Raphson direction and reduces the variance of the combined step. Second, we apply Levenberg damping, replacing $M_\theta$ with $M_\theta + \gamma I$ before inversion, and clamp the step in log-parameter space to a fixed maximum magnitude. Both modifications stabilise the iteration without biasing its target.

\subsection{Inference}
\subsubsection{Fixed Effect Standard Errors}
\citet{Louis1982} gives the marginal observed information for $\beta$ as
\begin{equation*}
    \mathcal{I}_\beta = \mathbb{E}_{\mathbf{u|\mathbf{y}}}\left[ - \frac{\partial^2 f(\mathbf{y|\mathbf{u},\beta)}}{\partial \beta \partial \beta^T}\right] - \text{Var}_{\mathbf{u|\mathbf{y}}}\left[ - \frac{\partial f(\mathbf{y|\mathbf{u},\beta)}}{\partial \beta}\right]
\end{equation*}
The conditional score and Hessian are:
\begin{equation*}
    \frac{\partial f(\mathbf{y|\mathbf{u},\beta)}}{\partial \beta} = X^T(\mathbf{y} - s(\mathbf{u})), \hspace{1cm} \frac{\partial^2 f(\mathbf{y|\mathbf{u},\beta)}}{\partial \beta \partial \beta^T} = X^T W(\mathbf{u}) X
\end{equation*}
where $s(\mathbf{u}) = \mathbb{E}[\mathbf{y}|\mathbf{u}]$. Substituting we get:
\begin{equation*}
    \mathcal{I}_\beta = X^T\mathbb{E}_{\mathbf{u|\mathbf{y}}}[W(\mathbf{u})]X - X^T\text{Var}_{\mathbf{u|\mathbf{y}}}[s(\mathbf{u})]X = \mathcal{B} - \mathcal{M}
\end{equation*}

Given our $K$ Monte Carlo samples $\mathbf{u}^{(k)}$ with self-normalised importance weights $w_k$ the matrices $\mathcal{B}$ and $\mathcal{M}$ are estimated as:
\begin{equation*}
    \hat{\mathcal{B}}_{MC} = X^T \text{diag}(\bar{W})X, \hspace{1cm} \bar{W}_i = \sum_{k=1}^K w_k W_{ii}(\mathbf{u}^{(k)})
\end{equation*}
and 
\begin{equation*}
    \hat{\mathcal{M}}_{MC} = \sum_{k=1}^K w_k \mathbf{g}_k \mathbf{g}_k^T, \hspace{1cm} \mathbf{g}_k = X^T\left(s(\mathbf{u}^{(k)}) - \bar{s} \right), \hspace{1cm} \bar{s} = \sum_{k=1}^K w_k s(\mathbf{u}^{(k)})
\end{equation*}
Then our marginal Monte Carlo information matrix is $\hat{\mathcal{I}}_\beta = \hat{\mathcal{B}}_{MC} - \hat{\mathcal{M}}_{MC}$

There is bias in the marginal variance for the Poisson-log model. The linearisation of $\exp(\eta)$ around the mode in a Laplace/PQL context underestimates the marginal mean by a factor related to the posterior variance \citet{Breslow1995}. In initial tests, the confidence intervals for $\beta_1$ showed under-coverage. When $\mathcal{B} - \mathcal{M}$ is estimated by Monte Carlo averaging over posterior samples, Jensen's inequality inflates the expected weights:
\begin{equation*}
    \mathbb{E}_{\mathbf{u}|\mathbf{y}}[\exp(\eta_i)] = \exp\left( x^T_i \beta + \bar{u}_i + \sigma^2_{i|y}/2 \right)
\end{equation*}
where $\sigma^2_{i|y} = \text{Var}(u_i | y)$ is the posterior marginal variance. This inflation is compounded across repeated datasets by the data-adaptivity of the posterior mean $\bar{\mathbf{u}}$, which covaries positively with $y$, contributing a further factor of $\exp((\sigma^2_i - \sigma^2_{i|y})/2)$ where $\sigma^2_i = (\Phi \text{diag}(\boldsymbol{\Lambda})\Phi^T)_{ii}$ is the prior variance. For the Poisson model, we therefore also consider a correction for the standard error by computing the information with the modified linear predictor $\tilde{\eta_i}^{(k)} = \eta_i^{(k)} - \sigma^2_i/2$, which yields effective weights
\begin{equation*}
    \tilde{W}_{ii} = \exp(x_i^T\beta + \bar{u}_i) \cdot \exp\left( - (\sigma^2_i - \sigma^2_{i|y})/2 \right) 
\end{equation*}
which retains the genuine spatial information through the posterior mean while removing Jensen-induced inflation. The correction is specific to unbounded link functions; for bounded links such as the logit, the weights are naturally capped and no correction is required.

\subsubsection{Random Effect Variance}
The joint variance of $(\beta,\mathbf{u})$ for HSGP is 
\begin{equation*}
    \begin{bmatrix}
        X^TW(\mathbf{u})X & X^TW(\mathbf{u})\Phi \\
        \Phi^TW(\mathbf{u})X & \Phi^TW(\mathbf{u})\Phi + \text{diag}(\boldsymbol{\Lambda}^{-1})
    \end{bmatrix}
\end{equation*}
We construct Wald intervals using the diagonal lower right component of the inverse of this matrix. 

For SPDE, the joint variance of $(\boldsymbol{\beta}, \mathbf{u})$ is
\begin{equation*}
    \begin{bmatrix}
        X^T W(\mathbf{u}) X & X^T W(\mathbf{u}) Z_A \\
        Z_A^T W(\mathbf{u}) X & Z_A^T W(\mathbf{u}) Z_A + Q
    \end{bmatrix}
\end{equation*}
Unlike the HSGP case, this matrix is sparse in its lower-right block and we avoid forming its inverse explicitly. Let $P = Z_A^T W Z_A + Q$ and let $L_P$ be the sparse Cholesky factor already cached from the posterior sampling step. The marginal variance of $\mathbf{u}$ is obtained by Schur complement against the $\boldsymbol{\beta}$ block, and the diagonal entries needed for Wald intervals at mesh vertices are extracted via Takahashi's selected-inverse recursion, which returns $[P^{-1}]_{jj}$ using only the sparsity pattern of $L_P$. Pointwise variances at observation or prediction locations $\mathbf{s}^*$ are then obtained by applying the corresponding row of the barycentric projector $A$.

\section{Stopping Criteria and Monte Carlo Sample Sizes}

\subsection{Stopping Criterion and Times}
In deterministic model fitting algorithms one can monitor the differences in the parameter estimates between successive iterations, and terminate the algorithm when the largest difference falls below some tolerance. However, Stochastic algorithms with constant step size though do not converge to a single value. The algorithm in this article is a form of stochastic gradient descent with approximately constant step sizes, which has a transient improvement phase, followed by a stationary, oscillating phase \citep{Murata1999}. Let $\hat{\mathcal{L}}^{(t)}$ be the log-likelihood on the $t$ step, and  $\Delta\hat{\mathcal{L}}^{(t)} = \hat{\mathcal{L}}^{(t)} - \hat{\mathcal{L}}^{(t-1)}$ be the differences. In the pre-convergence, improving phase $E[\Delta\hat{\mathcal{L}}^{(t)}] = \mu_{\Delta L}^{(t)} > 0$ and in the stationary phase we have $\mu_{\Delta L}^{(t)} = 0$. Our goal is to stop in the latter phase. 

Several authors have proposed stopping criteria for stochastic gradient descent algorithms \citep{pmlr-v84-chee18a,Patel2022}. These criteria need to balance the costs of a false negative, i.e. not stopping sufficiently soon and continuing with the algorithm at potentially high computational cost, and a false positive, i.e. stopping before the stationary phase. \citet{pmlr-v84-chee18a} propose montioring the cumulative sum of ratios of successive differences in parameter values, which they show becomes negative during the stationary phase. The procedure includes a burn-in phase to prevent premature termination. However, much of the literature in this area considers stochastic algorithms in the context of prediction, rather than estimation.

In the context of MCML algorithms, \citet{Caffo2005} proposed to effectively stop the algorithm when we reject a one-sided null hypothesis significance test $H_0:E[\Delta\hat{\mathcal{L}}^{(t)}] = 0$ versus $H_1: E[\Delta\hat{\mathcal{L}}^{(t)}] > 0$. Their approach monitors the upper bound of the confidence interval for $\Delta\hat{\mathcal{L}}^{(t)}$, which has estimated mean and standard error of $m^{(t)}_\Delta$ and $s^{(t)}_\Delta$, respectively. This approach has a high probability of stopping when the algorithm has converged, however, it is highly likely to require a large number of iterations post-convergence. If we choose to stop when $p<0.05$ then on average we require 20 post-convergence iterations. 

\begin{figure}
    \centering
    \includegraphics[width=0.8\linewidth]{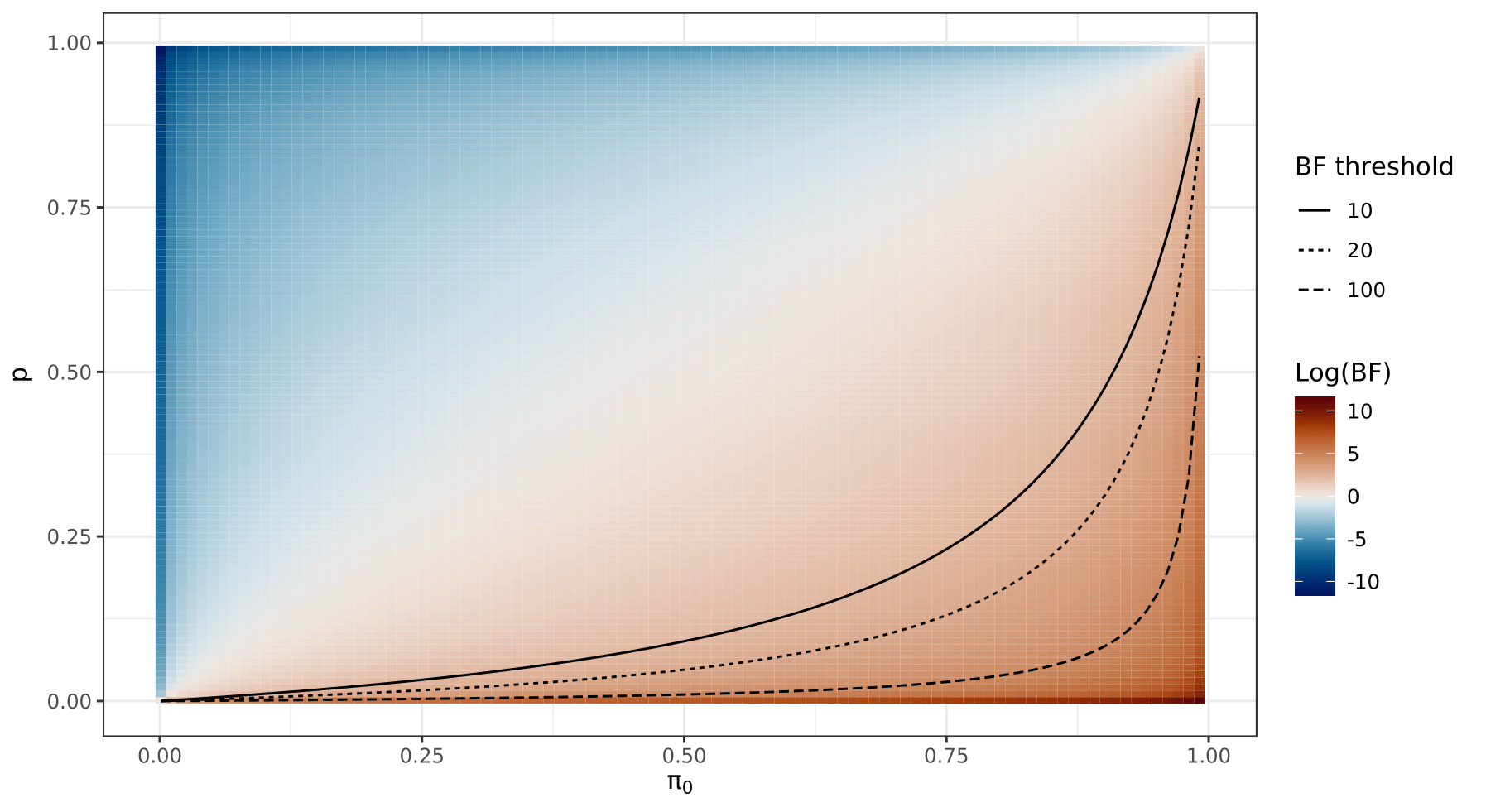}
    \caption{Values of the log Bayes Factor for different values of $\pi_0$ and $p$ with thresholds for different values of the Bayes Factor}
    \label{fig:bf}
\end{figure}

Our proposal is to incorporate the prior knowledge that the algorithm is more likely to have converged the more iterations that have passed. Consider the Bayes factor:
\begin{equation*}
    BF=\frac{Pr(\mu_{\Delta L}^{(t)} \leq 0 \vert \mathcal{D}^{(t)})}{Pr(\mu_{\Delta L}^{(t)} > 0 \vert \mathcal{D}^{(t)})} = \frac{f(\mathcal{D}^{(t)}\vert \mu_{\Delta L}^{(t)} \leq 0)}{f(\mathcal{D}^{(t)}\vert \mu_{\Delta L}^{(t)}> 0)}\frac{\pi_0}{1 - \pi_0}
\end{equation*}
where $\pi_0 = Pr(\mu_{\Delta L}^{(t)} \leq 0)$. Assuming the model $\Delta\hat{\mathcal{L}}^{(t)} \sim N(\mu_{\Delta L}^{(t)}, v_{\Delta L}^{(t)})$, then the likelihood ratio on the right-hand side is equivalent to $\frac{1-p^{(t)}}{p^{(t)}}$ where $p^{(t)}$ is the \textit{p}-value from a one sided test of the null $H_0:\mu_{\Delta L}^{(t)} = 0$ versus $H_1:\mu_{\Delta L}^{(t)} < 0$ \citep{Marsman2017}. We can then modify the $Pr(\mu_{\Delta L}^{(t)} \leq 0)$ prior probability as a function of $t$ and stop when the Bayes factor exceeds a predefined threshold. We specify the following model for the prior probability of convergence:
\begin{equation}
    Pr(\mu_{\Delta L}^{(t)} \leq 0) = 1 - \exp\left(- \left(\frac{t}{t_0}\right)^2 \right)
\end{equation}
which is approximately the cumulative distribution of a Weibull distribution with expected convergence time of $t_0$. Figure \ref{fig:bf} shows the Bayes Factors for different prior probabilities and \textit{p}-values. Even with high prior probabilities we still require relatively good evidence that the mean gradient difference is not positive, particularly for higher Bayes Factor thresholds to terminate the algorithm. 

For Newton-Raphson optimisation of a strongly convex function, gradient-based methods achieve linear convergence with rate $O(\kappa \log\frac{1}{\varepsilon})$ where $\kappa$ is the condition number of the Hessian matrix \citep{Nocedal2006,Bubeck2015}. The number of iterations to acheive convergence is approximately $\frac{\kappa}{2}\log(\frac{\vert \vert \beta^{(0)} - \beta_{ML} \vert \vert}{\varepsilon})$ to converge to a $\varepsilon$-ball around the maximum likelihood estimates $\beta_{ML}$. In the stochastic Newton-Raphson setting, we can only converge to within the Monte Carlo variance of the true maximum likelihood values. Therefore an approximate expected number of iterations to converge $t_0$ is:
\begin{equation*}
    t_0 \approx \frac{\kappa}{2} \log\left(\frac{\frac{1}{P}\sqrt{\vert \vert \beta^{(0)} - \beta_{ML} \vert \vert^2} }{\sqrt{\sigma^2_{MC}/\lambda_{min}m}} \right)
\end{equation*}
where $\kappa$ is the condition number of $M_\beta$ equal to $\lambda_{max}/\lambda_{min}$ which is the ratio of the largest to smallest eigenvalues of $M_\beta$, and where $\sigma^2_{MC}$ is a representative value of the Monte Carlo error. Evidently the convergence time depends on how far the starting values are from the maximum likelihood estimates. For the parameters $\beta$ we can start at the point estimates from the ordinary GLM model, however, for the covariance parameters we may need to use other strategies, such as an empirical variogram. Using the values of Monte Carlo error computed using the formulae described in the Supplementary Information, and starting values between 0.1 and 1.0 from the maximum likelihood estimates, we estimated values between 10 and 20 iterations for the Gaussian process models in the simulation study.

\section{Simulation Study}
We conduct a simulation study to estimate the Frequentist properties of the estimator of the HSGP model and to compare with INLA model fitting. We consider binomial and Poisson spatial Gaussian process models. We have implemented the algorithm in the package \textit{glmmrBase} (version 1.3.0) for the R programming language \citep{r_lang} (version 4.5.0). 

\subsection{Algorithms}
We compare several methods and algorithms. First, we consider HSGP with $m=5, 10, 15$ basis functions per dimensions and $L = 1.05, 1.20, 1.50, 2.00$, similar to \citet{RiutortMayol2023}. The number of Monte Carlo samples is either 1, 50, or 200. A Monte Carlo sample size of 1 corresponds to a Laplace Approximation as we use the posterior mean directly in place of Monte Carlo samples. Second, full maximum likelihood using the MCML algorithm described above and in the Supplementary Information with dense $D$. Third, INLA with a mesh size cut off of $\max(1/\sqrt{n}, 0.05)$ and using penalised complexity priors. 

\subsection{Data generating processes}
\subsubsection{Simulated Gaussian Processes}
For the models specified below, each observation $i$ is associated with a location $s_i \in A$, where $A$ is the unit square $[0,1] \times [0,1]$. We notate $y(s_i)$ as the outcome for observation $i$. The linear predictor at location $s_i$ is then:
\begin{equation}
\label{eq:linpred}
    \eta(s_{i}) = \beta_0 + \beta_1z_{i} + u(s_{i}) 
\end{equation}
where $z_i \sim N(0,1)$ is a covariate in the model. To facilitate comparison with INLA's built-in covariance functions we specify a Matern covariance function with shape $\nu = 1$:
\begin{equation*}
    \text{Cov}(\alpha(s_i), \alpha(s_{i'})) = \tau^2 \left(1 + \frac{\vert s_i - s_{i'}\vert}{\lambda}\right) \exp \left(-\frac{\vert s_i - s_{i'}\vert}{\lambda}\right)
\end{equation*}
where here $\vert . \vert$ is the Euclidean distance. 

We consider two data generating processes (DGP) for the Gaussian Process component. The ``smooth'' model with $\tau^2 = 1.0$ and $\lambda = 0.3$ and a ``rough'' model with $\tau^2 = 2.0$ and $\lambda = 0.15$. The shorter length scale may present more difficulty for HSGP with fewer basis functions.

\subsubsection{Poisson GLMM}
A common use of a spatial Poisson GLMM is fitting Log Gaussian Cox Process models, which are models of spatial point process data, like incident cases of a disease. A common computation approach to estimate the LGCP is to aggregate the cases to a regular lattice over the area of interest \citep{Taylor2013,Diggle2013}. The data for this model are therefore simulated on a regular grid covering the area and the locations $s_i$ are centroids of the grid cells. We note this is a somewhat artificial comparison for SPDE which does not require the computational grid for LGCP models \citep{Simpson2016}, but it provides a head-to-head comparison for HSGP. We consider grids of 10, 20, 30, and 40 cells per dimension. For the smooth DGP we use $\beta_0 = 0.0$ and $\beta_1 = 0.2$ and for the rough DGP $\beta_0 = -1.0$ and $\beta_1 = 0.2$.

\subsubsection{Binomial GLMM}
For the binomial GLMM we use the same linear predictor as (\ref{eq:linpred}) with a binomial-logit model $y(s_i) \sim Bin(N,\mu(s_i))$ and $\log(\mu(s_i)/(1-\mu(s_i)) = \eta(s_i)$. Data locations are sampled uniformly in the area of interest, and we consider sample sizes of $n = 100, 200, 400, 800, 1500$. For the smooth DGP we use $\beta_0 = 0.0$, $\beta_1 = 0.2$ and $N = 10$. For the rough DGP we use $\beta_0 = -1.0$, $\beta_1 = 0.2$ and $N = 1$.

\subsubsection{Evaluation}
For each scenario we extract the running time for model fitting (including mesh creation for INLA) and the point estimates and standard errors of $\beta$, $\theta$, and $\mathbf{u}$ (or equivalent posterior summaries for INLA). We estimate bias $\beta$ and for the covariance parameters, we report the median relative bias as a \%:
\begin{equation*}
    MRB(\hat{\tau^2}) = \left(\frac{\text{median}(\hat{\tau^2})}{\tau^2} - 1\right)*100
\end{equation*}
Both model fitting algorithms can enter pathological log-likelihood territory and the estimates can tend to positive or negative infinity, which can severaly affect the mean. We also report coverage of the 95\% confidence intervals (or posterior equivalent) for $\beta$ and $\mathbf{u}$.

\section{Results}
\subsection{Running Time}
HSGP was generally the fastest method at smaller sample sizes. For $m=10$ basis functions per dimensions fitting times for $n=1,000$ were around 1 second compared to around 3-5 seconds for SPDE approaches. The running time for HSGP is quadratic in the number of bases and approximately linear in the sample size. The mesh for the SPDE models was constrained in number of nodes above a sample size and became the faster method above approximately 2,000 observations at around 5-7 seconds. 

\subsection{Bias}
\subsubsection{Fixed Effects}
\begin{figure}
    \centering
    \includegraphics[width=\linewidth]{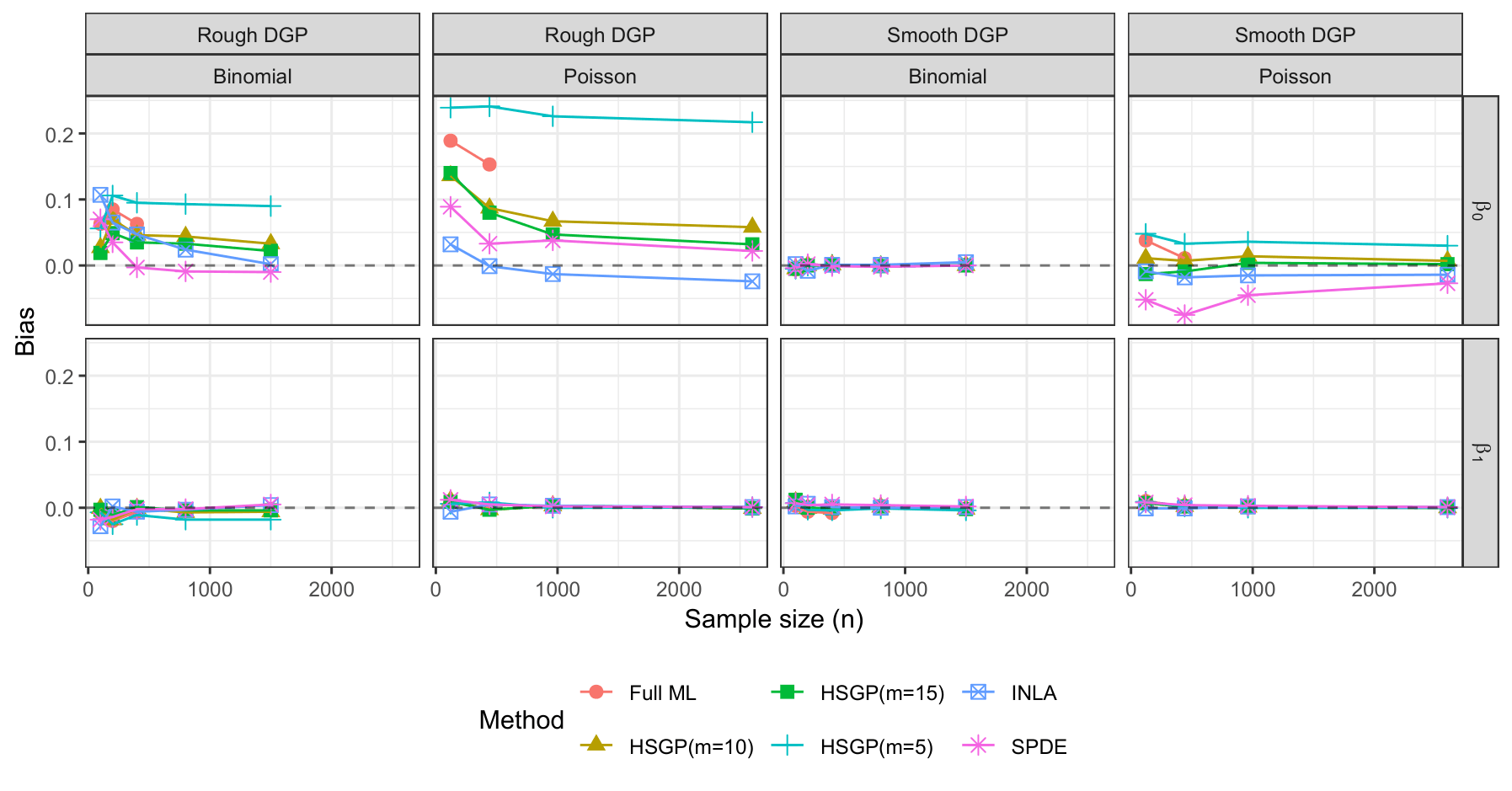}
    \caption{Absolute bias of the fixed effect parameters across}
    \label{fig:bias}
\end{figure}

For HSGP, we identified that $L=1.05$ produced the worst results for all $L$ values with performance also degrading for $L=2.00$. The results reported here are those for $L=1.20$, which generally exhibited the most favourable properties across all simulations. Figure \ref{fig:bias} reports the bias across the different algorithms. All methods were unbiased for $\beta_1$ across all sample sizes. The intercept exhibited more bias, especially for Poisson with HSGP methods, although the bias declined with sample size and sufficient basis functions suggesting consistency. The HSGP methods also had more bias for the rough DGP in the Poisson model than the smooth DGP, which also reduced with increased basis functions. SPDE methods were generally less biased for the intercept term, with INLA less biased than MCML for the Poisson model.

\subsubsection{Covariance parameters}
\begin{figure}
    \centering
    \includegraphics[width=\linewidth]{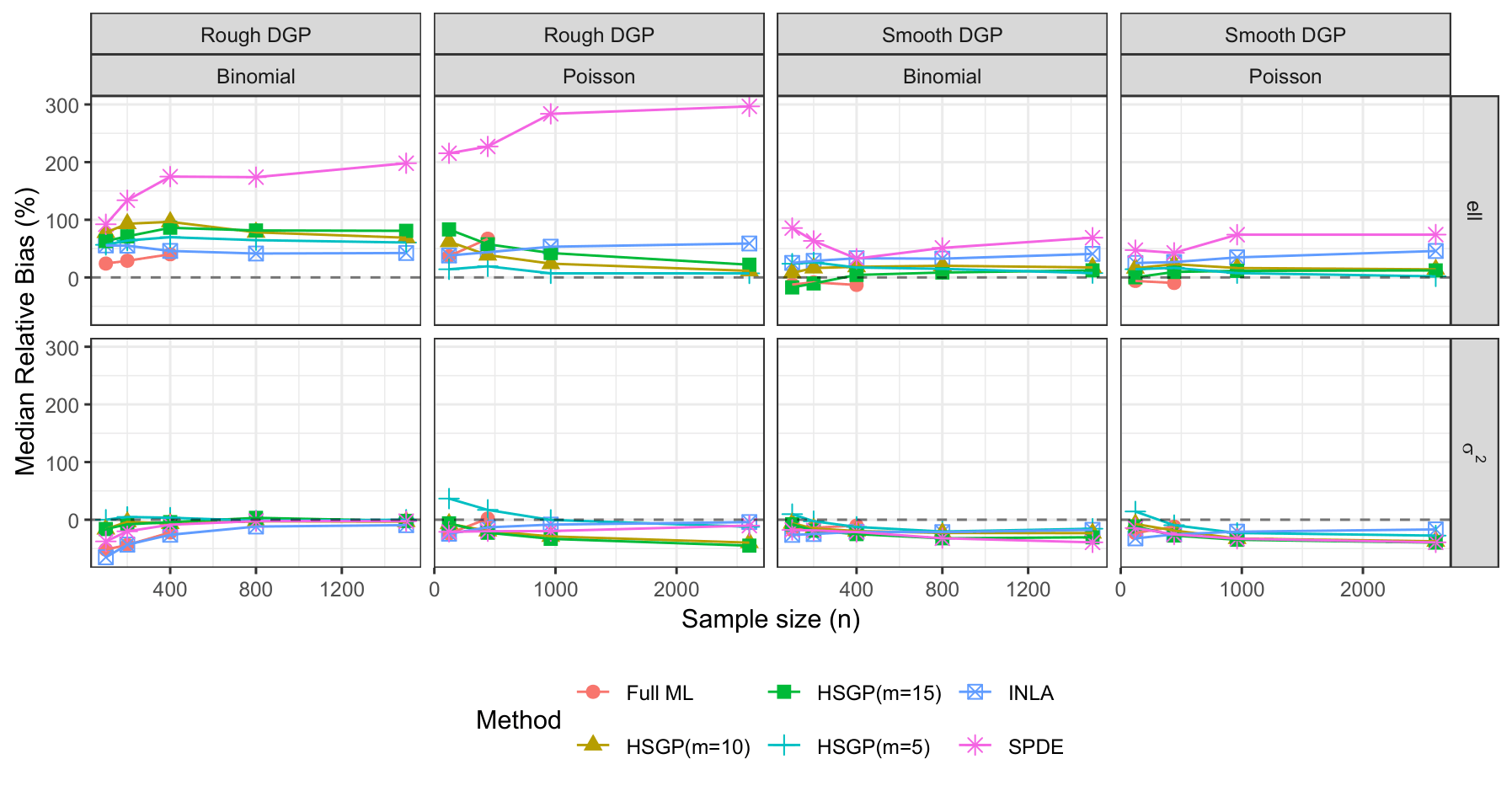}
    \caption{Median relative bias of the full ML, HSGP, SPDE, and INLA models for the covariance parameters $\tau^2$ and $\lambda$}
    \label{fig:mrb}
\end{figure}

Figure \ref{fig:mrb} shows the median relative bias of the covariance parameter estimates. The variance parameter $\tau^2$ was generally well identified and all estimators exhibited no or a small downward bias. The length scale parameter had an upward bias for all methods. The MCML SPDE model had the greatest upward bias of 50 - 300\%, particularly for the Poisson model and rough DGP, HSGP  methods with 10 or 15 basis functions per dimension had the smallest bias of 0 - 100\%, with INLA between the two except for the rough DGP and binomial model where at +50\% it had the smallest bias.

\subsection{Coverage}
\subsubsection{Fixed Effects}
\begin{figure}
    \centering
    \includegraphics[width=\linewidth]{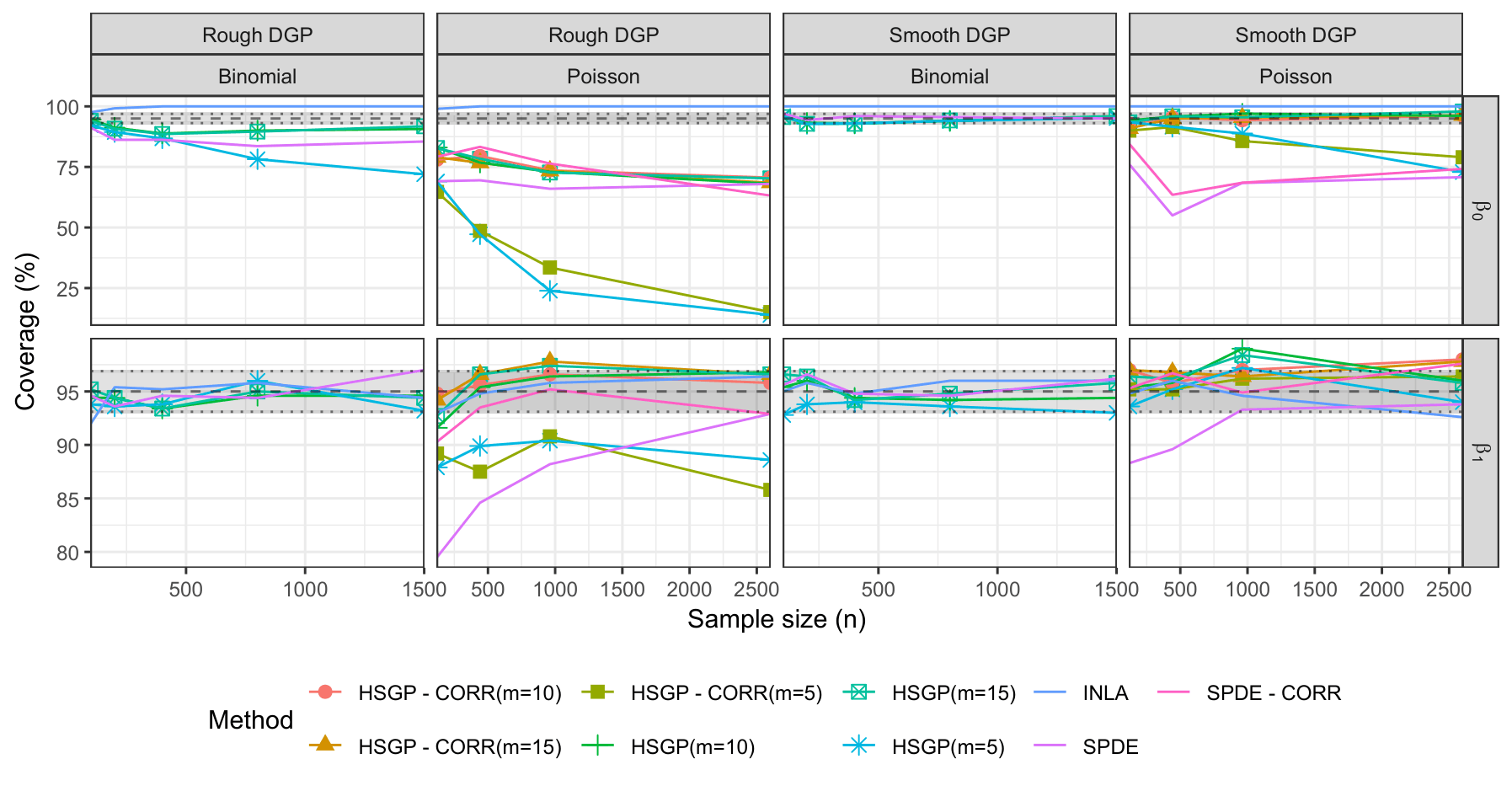}
    \caption{Coverage of the 95\% confidence intervals for full ML, HSGP, SPDE and corrected estimators (HSGP-CORR, SPDE-CORR) for Poisson, and 95\% credible intervals for INLA. The grey band is the 95\% confidence interval for the nominal value accounting for Monte Carlo error.}
    \label{fig:cover}
\end{figure}

Figure \ref{fig:cover} shows the estimated coverage of the 95\% confidence intervals. For $\beta_1$ and binomial, all methods produced nominal confidence intervals. For the Poisson model, the uncorrected MCML standard error estimator displayed high under-coverage for HSGP and SPDE but produced nominal or near-nominal intervals with the proposed correction. 

The intervals for the intercept parameter $\beta_0$ generally showed poor coverage, reflecting both the bias in Figure \ref{fig:bias} and standard error estimators. INLA's intervals were $\sim100\%$ for all models. For the smooth DGP HSGP had nominal coverage for both Binomial and Poisson (with the correction), while SPDE only had nominal coverage for Binomial. All MCML methods had significant undercoverage for the rough DGP.

\subsubsection{Random Effects}
\begin{figure}
    \centering
    \includegraphics[width=\linewidth]{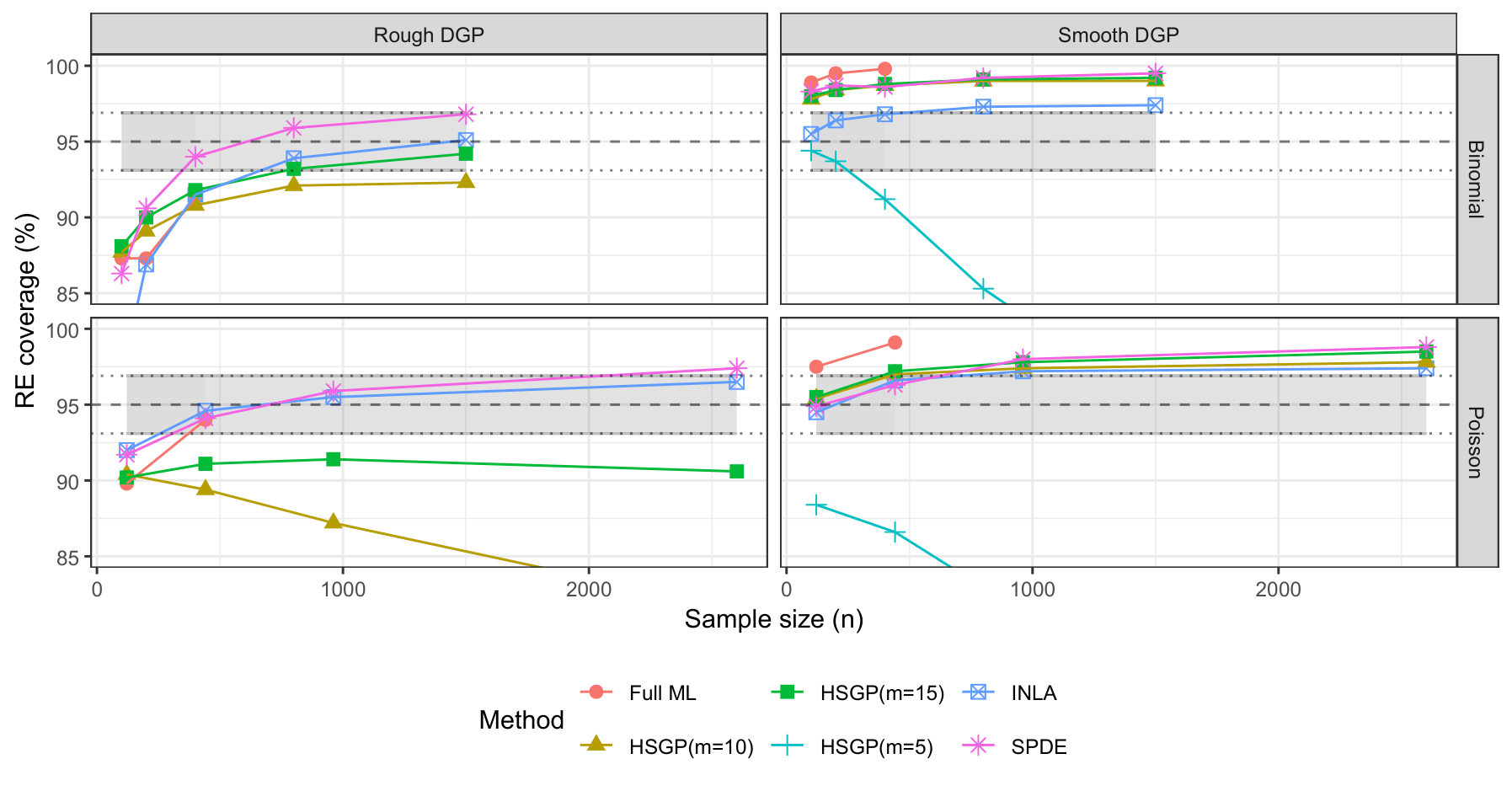}
    \caption{Coverage of the 95\% confidence interval or credible interval for the latent field estimates at the observed locations.}
    \label{fig:recover}
\end{figure}

Finally, Figure \ref{fig:recover} shows the coverage of 95\% prediction intervals for the random effects terms. No method consistently produced nominal intervals. For the smooth DGP intervals were nominal or moderately over-covered (95 - 98\%) with HSGP, SPDE, and INLA. For the rough DGP, HSGP consistently under-covered for Poisson (90 - 92 \%) with 15 basis functions per dimension. All methods produced prediction intervals with low coverage at smaller sample sizes with the rough DGP and binomial model, but achieved nominal coverage at large sample sizes. MCML with SPDE achieved nominal coverage at smaller sample sizes than other methods.

\section{Conclusion}\label{sec-conc}
We have demonstrated likelihood-based inference with a Gaussian process approximation can produce reliable and fast results for non-Gaussian spatial models. No method exhibited consistent unbiasedness and nominal intervals, however we draw several conclusions. For estimation of the effects of covariates, all methods produced an unbiased estimator with nominal confidence intervals. Our interest in the Frequentist properties of these estimators relates to growing interest in experimental design in spatial and geographic settings (e.g. \citet{Watson2025}), where control of type I error and coverage of confidence intervals is often a primary concern for estimator choice. For experiments, the parameter of interest is some type of treatment effect, which is uncorrelated with spatial effects by design and these approximations may be useful for spatial models in this context. For observational data, issues like spatial confounding \citep{Hodges2010} should be considered. Estimation of intercept or length scale parameters was generally unreliable, however.

Another key use of non-Gaussian spatial models is prediction and disease risk mapping. We reported the coverage of prediction intervals for the latent field. Here, the Frequentist characteristics of the HSGP MCML estimator was comparable to INLA except for the rough DGP and Poisson model, where HSGP intervals were moderately too narrow. There may be several explanations, such as a loss of variance from basis truncation. A potential correction may be use of Poisson-log normal distribution models \citep{Aitchson1989} accounting for the truncated basis variance. Future work may focus in this area. However, the SPDE approximation with MCML performed similarly to INLA and may be preferred, especially for ``rough'' process. In a non-Bayesian context, interval estimates can be obtained through several approaches, including profile likelihood methods or bootstrap procedures. Each of these approaches relies on different asymptotic approximations, with accuracy varying across model parameters and computational burden differing substantially between methods.

Our comparisons in this article are limited, and we only consider a small number of models. There may be certain specifications or parts of the parameter space where the algorithms struggle. Our Poisson model was artificial and does not represent the way in which one would typically estimate a Log Gaussian Cox Process model with SPDE. However, our aim was not to compare these algorithms across the whole space of mixed models, but to consider the feasibility of maximum likelihood algorithms in this area, which can be further adapted. Similarly, we only compared against an INLA estimator using a penalised complexity prior. However, our interest in INLA is predominantly as a popular computational method with reliable Frequentist properties for complex spatial data rather than a philosophical commmitment to Bayesianism. In this light, we demonstrate that SPDE can be used with maximum likelihood where non-Bayesian approaches are preferred.

We have demonstrated that approximate maximum likelihood estimation of mixed models with Gaussian Process models can be not only feasible, but fast and reliable. These methods may be a useful alternative when non-Bayesian methods are required. The algorithm can also be adapted to more complex models where it could provide improved inference, for example, we have adapted the algorithm for models of spatially aggregated point process data and non-stationary models where implementation of other methods may be complex. Further research may investigate performance in these settings.

\begin{appendices}

\section{Monte Carlo Error}
\subsection{Monte Carlo Error and Sample Sizes}
The number of Monte Carlo samples per iteration is also an important question. \citet{Caffo2005} suggested using adaptive sample sizes per iteration to improve convergence. One can set the sample size so that the Monte Carlo error in the parameter estimates is less than some proportion $p$ of the total variance including sampling variance. At convergence each Newton-Raphson iteration only adds Monte Carlo error as the gradient is zero. In the HSGP parameterisation, the random effects satisfy $\mathbf{v} \sim N(\mathbf{0}, \text{diag}(\boldsymbol{\Lambda}(\boldsymbol{\theta})))$ with linear predictor $\boldsymbol{\eta} = X\boldsymbol{\beta} + \Phi\mathbf{v}$, so the prior precision $D^{-1} = \text{diag}(\boldsymbol{\Lambda}^{-1})$ is diagonal and the covariance derivatives reduce to $\partial D / \partial \theta_i = \text{diag}(\boldsymbol{\Lambda}'_i)$, where $\boldsymbol{\Lambda}'_i = \partial \boldsymbol{\Lambda} / \partial \theta_i$. Let $H = \Phi^\top W \Phi + \text{diag}(\boldsymbol{\Lambda}^{-1})$ denote the $M \times M$ posterior precision in $\mathbf{u}$-space, and define $\mathbf{q}_i = \boldsymbol{\Lambda}'_i / \boldsymbol{\Lambda}^2$ componentwise.

The variance of the Monte Carlo error for the covariance parameters is approximately $M_{\theta,MC} = M_\theta^{-1} V_\theta M_\theta^{-1}$, where $V_\theta$ is a diagonal matrix with $ii$th entry:
\begin{align*}
    V_{\theta,ii} &= \sum_{k=1}^M q_{i,k}^2 (H^{-1})_{kk} + 2 (\boldsymbol{\mu}_u \odot \mathbf{q}_i)^\top H^{-1} (\boldsymbol{\mu}_u \odot \mathbf{q}_i)
\end{align*}
where $\odot$ denotes the Hadamard product and $\boldsymbol{\mu}_u$ is the current posterior mean of $\mathbf{u}$. For the parameters $\beta$ we similarly calculate $M_{\beta,MC} = \mathcal{I}_\beta^{-1} V_\beta \mathcal{I}_\beta^{-1}$ with:
\begin{align*}
    V_\beta &= X^\top W(\mathbf{u}) A \, H^{-1} \, A^\top W(\mathbf{u}) X
\end{align*}
The number of Monte Carlo samples to maintain the proportion of Monte Carlo variance to proportion $p$ for the $i$th parameter $\theta$ is:
\begin{equation*}
    m = \frac{1-p}{p}\frac{M_{\theta,MC,ii}}{M_{\theta,ii}}
\end{equation*}
and equivalently for the parameters $\beta$. We can then select the largest value over all the parameters. The parameters are unknown prior to model fitting and so reasonable guesses can be used to identify an initial sample size value. The number of samples could be updated iteratively as the parameters are updated.

\section{Full maximum likelihood specification}
\subsection{Sampling random effects}
Our target distribution for the samples of random effects is the posterior density $f_{\mathbf{u}|\mathbf{y}}(\mathbf{u}|\mathbf{y},\hat{\beta}^{(t)},\hat{\theta}^{(t)}) \propto f_{\mathbf{y}|\mathbf{u}}(\mathbf{y}|\mathbf{u},\hat{\beta}^{(t)})f_{\mathbf{u}}(\mathbf{u}|\hat{\theta}^{(t)})$. Both the first and third steps of the algorithm make use of the Cholesky decomposition of $D = LL^T$, which only needs to be calculated once per iteration. Thus, to avoid the inversion of $D$, instead of generating samples of $\mathbf{u}$ directly, we instead sample $\mathbf{v}$ from a proposal distribution based on the model:
\begin{align}
\begin{split}
\label{eq:mcmc}
    \mathbf{y} &\sim G(h(X\hat{\beta} + ZL\mathbf{v}); \hat{\phi}) \\
    \mathbf{v} &\sim N(0,I)
    \end{split}
\end{align}
The approximate posterior mean, equivalent to the posterior mode in the Laplacian approximation, of the random effects is:
\begin{equation}
\label{eq:v}
    \bar{\mathbf{v}}^{(t+1)} = (L^TZ^TW(\boldsymbol{\beta}^{(t)},\bar{\mathbf{v}}^{(t)})ZL+I)^{-1}(Z^TW(\boldsymbol{\beta}^{(t)},\bar{\mathbf{v}}^{(t)})ZL\bar{\mathbf{v}}^{(t)} + Z^T(y - \mu(\boldsymbol{\beta},\bar{\mathbf{v}}^{(t)})))
\end{equation}
where $W = \text{diag}\left( \left(\frac{\partial h^{-1}(\boldsymbol{\eta})}{\partial \boldsymbol{\eta}}\right)^2 \text{Var}(\mathbf{y}| \mathbf{u})\right)^{-1}$, which are recognisable as the GLM iterated weights \citep{breslow1993approximate, mccullagh2019generalized}. As both $W$ and $\mu$ depend on $\bar{\mathbf{v}}^{(t+1)}$, we iteratatively update the mean until convergence as an iteratively re-weighted least-squares approach. We then draw samples from the proposal distribution $\mathbf{v}^{(t+1)} \sim N(\bar{\mathbf{v}}^{(t+1)}, (L^TZ^TWZL+I)^{-1})$ and transform to $\mathbf{u}^{(t+1)} = L\mathbf{v}^{(t+1)}$.

We generate importance sampling weights for each Monte Carlo sample:
\begin{align}
\begin{split}
    w_{k}^{*(t)} &= \exp\left(\log f_{\mathbf{y}|\mathbf{u}}(\mathbf{y}|\mathbf{u}^{(k,t)},\boldsymbol{\beta}^{(t)},\phi^{(t)}) +  \log(f_{\mathbf{u}}(\mathbf{u}^{(k,t)}|\theta)) - \log (h(\mathbf{v}^{(k,t)}\vert\boldsymbol{\mu}_{\boldsymbol{v}}^{(t)},V_\mathbf{v})) \right) \\    
    &\propto \exp\left( \log f_{\mathbf{y}|\mathbf{u}}(\mathbf{y}|\mathbf{u}^{(k,t)},\boldsymbol{\beta}^{(t)},\phi^{(t)}) - 0.5  \mathbf{u}^{(k,t)T}D^{-1}\mathbf{u}^{(k,t)} + 0.5\mathbf{v}^{(k,t)T}V_\mathbf{v}\mathbf{v}^{(k,t)} \right) \\
    w_{k}^{(t)} &= \frac{w_{k}^{*(t)}}{\sum_{l=1}^m w_{k}^{*(t)}}
    \end{split}
\end{align}

\subsection{Covariance parameter estimation}
The multivariate Gaussian density is:
\begin{align}
\begin{split}
\label{eq:theta_ll}
    \log f_{\mathbf{u}}(\mathbf{u}|\theta) &= -\frac{m}{2}\log(2\pi) - \frac{1}{2}\log \text{det}(D) - \frac{1}{2}\mathbf{u}^T D^{-1} \mathbf{u}
\end{split}
\end{align}
\citet{mcculloch1997maximum} suggested maximising the likelihood function directly to estimate $\theta$ and we are not aware of any alternative approaches to fitting the covariance parameters in the literature. However, a Newton-Raphson step can also be used here. The gradient of the log-likelihood with respect to $\theta$ is:
\begin{align*}
   \frac{\partial \log f_{\mathbf{u}}(\mathbf{u}|\theta)}{\partial \theta_i} &= -\frac{1}{2} \text{trace}\left( D^{-1} \frac{\partial D}{\partial \theta_i} \right)  + \frac{1}{2} \text{trace}\left( D^{-1} \mathbf{u} \mathbf{u}^T D^{-1} \frac{\partial D}{\partial \theta_i}\right)
\end{align*}
and the negative Hessian, i.e. the inverse variance matrix, is the matrix $M_\theta$ with elements \citep{Stroup2012}:
\begin{equation}
\label{eq:hess_cov}
    M_{\theta_{ij}} = -\frac{1}{2}\text{trace}\left( D^{-1} \frac{\partial D}{\partial \theta_i} D^{-1} \frac{\partial D}{\partial \theta_j}  \right) + \text{trace}\left( D^{-1} \mathbf{u}\mathbf{u}^T D^{-1} \frac{\partial D}{\partial \theta_i} D^{-1} \frac{\partial D}{\partial \theta_j}  \right)
\end{equation}
We then update the parameters as:
    \begin{equation*}
        \theta^{i+1} = \theta^i + E_u[M_\theta^{-1}] E_u\left[ \frac{\partial \log f_{\mathbf{u}}(\mathbf{u}|\theta)}{\partial \theta_i} \right]
    \end{equation*}
where the expected value is estimated using the Monte Carlo samples:
\begin{equation*}
    E_u\left[ \frac{\partial \log f_{\mathbf{u}}(\mathbf{u}|\theta)}{\partial \theta_i} \right] \approx -\frac{1}{2} \text{trace}\left( D^{(t)-1} \frac{\partial D^{(t)}}{\partial \theta_i} \right)  + \frac{1}{2} \sum_{k=1}^{m_t} w_k^{(t)}    \mathbf{u}^{(k,t)T} D^{(t)-1} \frac{\partial D^{(t)}}{\partial \theta_i}D^{(t)-1}\mathbf{u}^{(k,t)}
\end{equation*}
and similarly for the Hessian matrix.




\end{appendices}


\bibliography{smaxlik}% common bib file

\end{document}